\documentclass[sigconf]{acmart}

\copyrightyear{2022}
\acmYear{2022}
\setcopyright{acmcopyright}
\acmConference[ICSE '22]{44th International Conference on Software Engineering}{May 21--29, 2022}{Pittsburgh, PA, USA}
\acmBooktitle{44th International Conference on Software Engineering (ICSE '22), May 21--29, 2022, Pittsburgh, PA, USA}
\acmPrice{15.00}
\acmDOI{10.1145/3510003.3510627}
\acmISBN{978-1-4503-9221-1/22/05}


\usepackage[labelformat=simple]{subcaption}

\usepackage{multirow}
\usepackage{listings}
\usepackage{algorithm}
\usepackage{algpseudocode}
\usepackage{amsmath}
\usepackage{xcolor}
\usepackage{xspace}
\usepackage{array}
\usepackage{appendix}
\usepackage{multirow}
\usepackage{makecell}
\usepackage{arydshln}
\usepackage{enumitem}
\usepackage{graphicx}
\usepackage{dashrule}
\usepackage{float}
\usepackage{pifont}
\usepackage{colortbl}
\usepackage{amsthm}

\usepackage{tabulary}
\usepackage{tcolorbox}
\usepackage{threeparttable}
\usepackage[misc]{ifsym} 

\newcommand{\tool}{\textsc{ModX}\xspace}

\begin{document}
\title{ModX: Binary Level Partially Imported Third-Party Library Detection via Program Modularization and Semantic Matching}

\author{Can Yang$^{1,2}$, Zhengzi Xu$^{*3}$, Hongxu Chen$^{4}$, \
Yang Liu$^{3}$, Xiaorui Gong$^{1,2}$, Baoxu Liu$^{1,2}$}
\email{yangcan@iie.ac.cn, zhengzi.xu@ntu.edu.sg, chenhongxu5@huawei.com}
\email{yangliu@ntu.edu.sg, gongxiaorui@iie.ac.cn, liubaoxu@iie.ac.cn}

\affiliation{
    \institution{School of Cyber Security, UCAS$^1$; Institute of Information Engineering, CAS$^2$; School of Computer Science and Engineering, NTU$^{3}$; Huawei Technologies Co., Ltd.$^{4}$}
}

\begin{abstract}
With the rapid growth of software, using third-party libraries (TPLs) has become increasingly popular.
The prosperity of the library usage has provided the software engineers with a handful of methods to facilitate and boost the program development.
Unfortunately, it also poses great challenges as it becomes much more difficult to manage the large volume of libraries.
Researches and studies have been proposed to detect and understand the TPLs in the software. 
However, most existing approaches rely on syntactic features, which are not robust when these features are changed or deliberately hidden by the adversarial parties. 
Moreover, these approaches typically model each of the imported libraries as a whole, therefore, cannot be applied to scenarios where the host software only partially uses the library code segments.

To detect both fully and partially imported TPLs at the semantic level, we propose \tool, a framework that leverages novel program modularization techniques to decompose the program into fine-grained functionality-based modules. 
By extracting both syntactic and semantic features, it measures the distance between modules to detect similar library module reuse in the program. 
Experimental results show that \tool outperforms other modularization tools by distinguishing more coherent program modules with 353\% higher module quality scores and beats other TPL detection tools with on average 17\% better in precision and 8\% better in recall.

\end{abstract}


\maketitle

\newcommand\blfootnote[1]{%
\begingroup 
\renewcommand\thefootnote{}\footnote{#1}%
\addtocounter{footnote}{-1}%
\endgroup 
}
\blfootnote{* corresponding author.}

\section{Introduction}\label{sec:intro}
With the rapid development of commercial software, third-party library (TPL) reuse has become more and more popular to ensure high program quality and reduce the unnecessary development costs.
According to~\cite{gartner}, over 90\% of organizations leverage TPLs in application development. Both GitHub~\cite{github_annual_report_2020} and 
Sonatype~\cite{sonatype_supply_chain_report_2019} report that over 80\% of most applications’ code comes from library dependencies.
However, as the size of the software grows bigger and more libraries with different dependencies are involved, it is difficult to track all the imported TPLs accurately. 
The massive use of the uncontrolled libraries will result in issues in the areas such as code auditing (licence violations)~\cite{mancoridis1999bunch, duan2017osspolice, zhan2020automated, yuan2019b2sfinder}, malware affection~\cite{hamlen2019automated}, and unexpected vulnerability introduction~\cite{eghanrecovering}.
Understanding which libraries have been imported has become the key to address these issues. As a result, TPL detection works have been proposed, which extract features from known libraries and match them in the target software.   
For example, BAT~\cite{hemel2011finding} searches the reliable constants and strings in the program to detect TPLs.
OssPolice~\cite{duan2017osspolice} also leverages the invariant literals to detect TPLs with a hierarchical indexing scheme.
Moreover, works~\cite{ma2016libradar, li2017libd, zhan2020automated, zhang2019libid} have been proposed to improve the TPL detection ability on Android applications with package dependency identification.

However, existing feature matching-based approaches have two limitations. 
First, they embed features from the entire TPLs. 
If the program only imports part of the library, the detection algorithm may fail due to the lack of fully matched features. 
To detect the partially imported libraries, one possible solution is to match the library at a more fine-grained level. 
The only existing ready-to-use fine-grained unit in the program is the function. 
Methods~\cite{Ding2019Asm2Vec, gemini2017, zuo48neural} have been proposed to match the similar functions between the programs and libraries to detect the TPL usage. 
However, the matching algorithms are not robust at binary level.
It is because the functions are very likely to be changed due to different compiler settings~\cite{Ding2019Asm2Vec}. 
Therefore, choosing a matching unit which is not subject to change becomes the key in partial library detection.

The program module, as a conceptual unit, fits this need well due to the following reasons. 
First, it consists of several functions which are combined together to achieve a common functionality. 
Since the program reuses the library by importing the functionality groups, the module can be regarded as the basic fine-grained unit.
Second, since within a module, the functions are connected to each other to form a call graph, the module itself will be enriched with more semantic graphical features, which are unlikely to be changed by compilation. 
It helps to make the module matching more accurate and robust in the practical real-world TPL detection.
However, to our best knowledge, there are only few works on binary level program modularization.
BCD~\cite{karande2018bcd} is the state-of-the-art static approach to decompose the binary executables into modules. 
However, the modules it generated usually contain isolated functions, which will hinder the TPL detection in the later step. 
Therefore, the \textbf{first challenge} of this work is  to divide the given program into meaningful and practical modules.

The second limitation of the existing works is that they rely too much on syntactic features, especially the strings, to detect TPLs, since strings often bring direct indication of the library information.
However, this kind of features may be deliberately modified by others to hide the library information~\cite{biondi2018tutorial_malware_evasion}. 
Especially within modern malware, strings obfuscation has been one of the most commonly used evasion techniques~\cite{chakkaravarthy2019survey_malware}.
To overcome the drawbacks of using pure syntactic features, plenty of function matching and code clone detection researches~\cite{zuo48neural, Ding2019Asm2Vec, 2020DeepBinDiff, gemini2017, BinSim, chandramohan2016bingo, eschweiler2016discovre} have been proposed to embrace more semantic features.
However, these works focus on function level features, which may not be accurate in measuring module similarity. Also, the module possesses unique features that can help to distinguish themselves which are not captured by existing works.
Thus, the \textbf{second challenge} of this work is to accurately measure the semantic level similarity between the modules by extracting suitable features.

To this end, we propose \tool, a framework, which utilizes a novel modularization technique to decompose the program and library into modules and to semantically match them to detect either fully or partially imported TPLs. 
Inspired by the community detection algorithms~\cite{newman2004finding,newman2004fast, blondel2008fast, arenas2007size}, firstly, \tool defines the module quality score to assess the coherence of the function clusters. Then, for a given program or a library, it starts to group individual functions to form modules while maximizing the overall module quality score. After the programs and libraries have been modularized, \tool extracts both syntactic and semantic features from inter- and intra-module levels and measures the similarity between the modules. Based on the similarity, \tool will match and detect the 
presence of library modules in the program so that it can find the fully/partially imported TPLs. 
The experimental results show that \tool achieves 90.1\% precision and 78.2\% recall in TPL detection of self-crafted programs and 84.3\% precision and 61.7\% recall in real-world software, which outperforms other TPL detection tools.
Moreover, since the modularization algorithm is a stand-alone technique, it also has great applicants besides TPL detection. 
We also test its possibilities in different software engineering tasks such as reverse engineering and attack surface detection.

In summary, our main contributions are as follows:
\begin{itemize}
\item We propose a binary level program modularization algorithm to decompose a program into functionality-based modules, and develop metrics to assess the module quality.
\item We propose a semantic measurement algorithm to calculate the similarities between modules. 
\item We conduct TPL detection experiments on 128 real-world projects, in which \tool outperforms the state-of-the-art tools over 17\% in accuracy on average.
\item We evaluate the potential applications of the program modularization algorithm, such as reverse engineering and attack surface detection. 
\end{itemize} 

\section{Background}\label{sec:preliminary}
\subsection{Motivating Example}
In this section we illustrate our motivation with a real-world example. 
\textit{Watcher}~\cite{mal_motivation} is a malware used as a secret implant for monitoring network traffics.
We collect and upload the binary of \textit{Watcher} variant to the online platform VirusTotal~\cite{virustotal1}, which performs malware detection via 60 anti-virus engines. The result shows that only 7 out of 60 leading security vendors successfully detect the malware~\cite{virustotal}. 
The rest fail to detect the malware variant because it changes the binary instructions and the string literals to obfuscate itself.

To precisely detect the malware, security experts can use component analysis to determine the TPLs used by this malware as an indicator of the malware presence.
However, after the malware has been detected and its signature has been recorded in the anti-virus database, \textit{Watcher} also starts to evolve and hide itself. 
It removes all the strings inside the program since it does not need them to carry malicious activities.
Also, instead of using the entire pcap library or dynamically linking it, it only uses 8 export functions (The entire pcap library has 84 export functions). 
However, after the evolution, existing tools fail to find the library. 
According to our experiment, the state-of-the-art TPL detection tool BAT~\cite{hemel2011finding} outputs several false positives. 
Thus, the malware successfully hides the pcap library and escapes from the anti-malware detection. 

We propose the program modularization technique to divide the pcap library into 16 modules. 
We match the modules in the malware binary and detect that it reuses 3 of the modules. 
Therefore, we have provided a strong evidence to confirm the binary to be Watcher. 
The approach is more robust since the malware cannot live without the support of pcap. No matter what changes the malware makes to hide the library, as long as it does not change the function semantics, our tool can still find the trace of the library pcap.

\subsection{Background Information}
In this section, we briefly discuss about some software engineering concepts used in our paper.

\subsubsection{Third-Party Library}
TPL is a reusable software component being developed by some parties other than the original development vendor. It is distributed freely or under certain licence policies. 
It is used to avoid the repeating development of software with the same functionalities so that it can save time and resources.
However, due to lack of support from the third parties, using it also introduce dependency issues and security concerns. 

\subsubsection{Community Detection Algorithm} In a complex relation network, nodes tend to be gathered to form community structures. The community detection algorithm aims to reveal the hidden grouping information of the communities, which are frequently used in distributed network systems. It partitions the network graph into small clusters and detects the communities. 
In this work, the entire program or library can be regarded as a graph network with the functions representing the nodes. Program modularization is similar to the community detection algorithm, which tries to group functions into different communities (modules).

\subsubsection{Binary Code Clone Detection}
Binary code clone detection tries to find similar functions in the binary executables.
It is often used to audit the software originality and to search for recurring software bugs caused by code reuse and sharing.
The traditional algorithms extract different features to represent the code and measure the code similarity based on these features.
In this work, we aim to propose algorithms to measure the similarity between modules rather than functions so that it can be more robust to detect TPLs. We follow a similar approach as the traditional clone detection but with a different feature set.

\section{methodology}\label{sec:method}
\subsection{Overview}

\begin{figure*}[t]
\centering
\includegraphics[width=\linewidth]{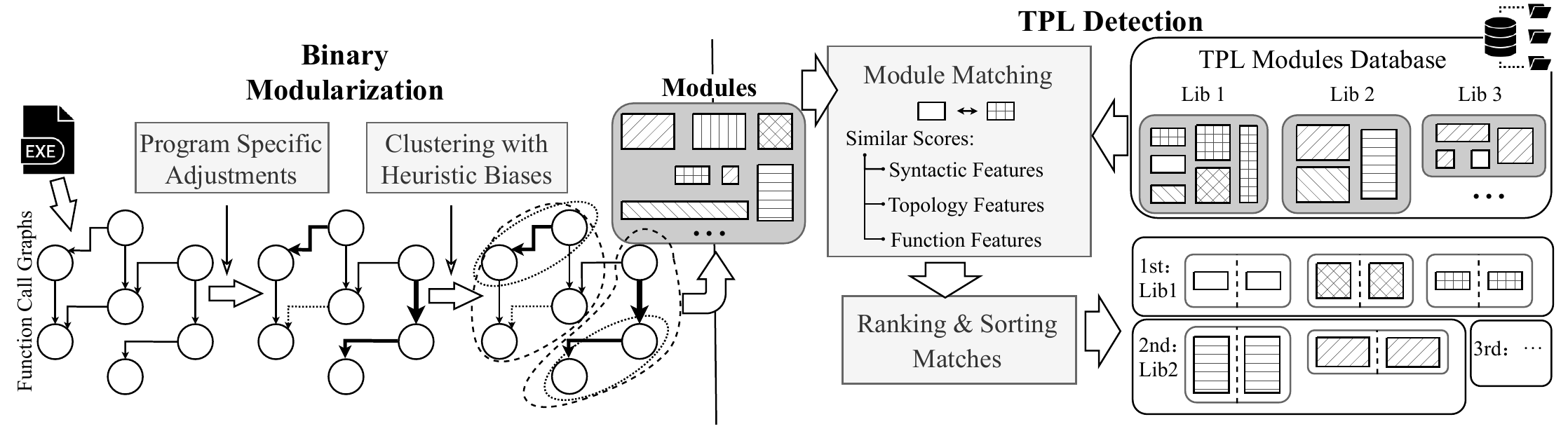}
\caption{Overall Workflow of \tool}
\label{fig:Workflow}
\vspace{-5pt}
\end{figure*}

Figure~\ref{fig:Workflow} shows the workflow of \tool. It consists of two phases, namely Binary Modularization and TPL Detection, to predict TPLs from a binary program. 
In the first phase, it proposes a module quality metric, which is based on community detection algorithm with program specific adjustments.
Then, it leverages a novel algorithm with heuristic biases to decompose the binary into modules based on the metric. 
In the second phase, \tool performs the TPL detection by matching program modules with TPL modules.
It extracts syntactic features, graph topology features, and function level features to measure the similarity between modules. 
After the matching, it also introduces module and library importance scores to help improve the library detection accuracy.

\subsubsection{Assumptions}
First, in this work, we assume that the modules of the program do not overlap with each other. For example, if module $A$ and $B$ both call the function $f$, then $f$ will have a high chance to be divided into a separated module $C$. $f$ will not belong to either $A$ or $B$.
Second, we assume that the content of each TPL will not change significantly. Since \tool aims to match TPLs across different versions using semantic features, if the semantics of the library have been changed significantly, \tool will fail to produce accurate results.

\subsection{Binary Program Modularization}
In our paper, the program modularization technique consists of two components, the module quality metric and the actual modularization algorithm.
The module metric aims to measure the quality gain from grouping functions into clusters, and the modularization algorithm combines the functions in the way which will maximize the overall module quality score. 

\subsubsection{Module Quality Assessment Design}
The program consists of functions which are connected with each other through function calls. The relationships can be represented by a call graph with functions as the nodes and calls as the edges. 
Functions with similar functionalities are likely to appear close to each other to form a community in the graph. 
The program modularization process aims to find these communities, which is very similar to the community detection in a network.
Therefore, to design a sound and practical module quality assessment metric, we adopt the community detection quality metrics as the baseline. 
Then, we modify the metrics with software specific heuristics to fit in the specific program modularization task.

\noindent
\textbf{Girvan–Newman Algorithm}
Inspired by the community detection algorithm, we choose Girvan–Newman Modularity Quality (GN-MQ)~\cite{newman2004finding} as the baseline metric since it has a good approximation on the program structure. 
It is the first algorithm proposed for modularity optimization, and has far-reaching impacts on following researches~\cite{blondel2008fast, arenas2007size, khan2017network}.
Basically, given a network which has been divided into multiple clusters, the metric counts the connected edges between each pair of nodes from the same clusters and sums the number of such occurrences with adaptive weights based on node degrees.
If there is no connection between the nodes in the same cluster, the weight will be assigned with negative values, which decreases the overall quality score. 
Specifically, the quality is calculated according to the Equation~\ref{eq_1} 

\begin{equation} 
    Q = \frac {1}{2m} \sum_{i,j}[A_{ij} - \frac{k_i k_j}{2m}]\delta(C_i, C_j) \label{eq_1}
    \end{equation}

\noindent
where $i$ and $j$ denotes the $i$th node and the $j$th node in the graph respectively, $A_{ij}$ denotes whether node $i$ and $j$ are connected or not, which has a value either 1 or 0, $k_i$ denotes the in- and out- degree of node $i$, $m$ is the number of edges in the graph, $C_i$ is the community where node $i$ belongs to, $\delta(C_i,C_j)$ stands for whether node $i$ and $j$ belong to the same cluster, which has a value either 1 or 0. 
As shown in this Equation, if the nodes $i$ and $j$ belong to the same cluster and they are connected to each other, then the quality score will increase. Otherwise, if the two nodes from the same cluster are not connected, the score will be decreased since $A_{ij}$ will be set to 0 and the term $A_{ij} - k_ik_j/2m$ will become negative. Therefore, in this metric, the high quality score reflects that the high coherence among the nodes within the cluster. Moreover, due to the negative term $-k_ik_j/2m$, nodes having less in- and out-degree will have more weights than others. Therefore, the metric also discourages the connectivity between nodes from different clusters.

\noindent
\textbf{Function Volume Adjustment.}
Besides the connectivity between nodes, the program modules have unique features that can be used as the module quality indicators. Function volume is one of them, which is specified by the number of statements in the function. 
In the program, functions that have large volumes tend to perform some core functionalities, whereas, small functions will likely be the utility functions~\cite{gemini2017, andriesse2016indepth}. 
A complete and coherent program module will consist of a small group of large-volume functions to perform the core functionalities and some small-volume functions, which are around the core group to provide useful utilities. 
Therefore, we propose the function volume weight propagation algorithm to add the weight adjustment to the metric so that it can favour the complete and coherence modules. 

The aim of the propagation algorithm is to assign different weights to each of the functions based on its volume and connectivity. 
It functions in a way that is similar to the PageRank~\cite{page1999pagerank} algorithm in website ranking.
For programs that have hierarchical structures, the functions at the top levels tend to control the behaviour of the low-level functions via function calls. 
The propagation algorithm guarantees that the top-level functions will receive more attention compared to the low-level ones, which results in more weights being assigned to the top-level functions. 
Therefore, when we modularize the programs, we are able to begin with these heavy-weighted functions to avoid generating modules with only small utility functions.

The detailed steps are as follow:
First, each function is initialized with its own volume value (e.g. the number of statements).
Then, we check the out-degree of each function and look for the end node which has 0 out-degree. Since the end node does not call other functions, its weight will not be affected by the rest of functions in the propagation.
Next, the weight of the end node will be propagated backward to its parent node (the caller function). We identify the number of function calls in the parent and adjust the weights by normalizing them against the number of calls. The propagation is defined as Equation~\ref{eq_fv},
\begin{equation} 
    FV^{\prime}(u) = FV(u) + c\sum_{v \in E(u)}{\frac{FV(v)}{C_v}} \label{eq_fv} 
\end{equation}
\noindent
where $FV$ refers to the function volume weight, $u$ and $v$ represent the function nodes with $u$ calls $v$. $E(u)$ is the set of the end nodes, which $u$ calls. $C_v$ denotes the number of caller functions of $v$. $c$ is a factor used for normalization. The $FV$ of the top level node $u$ will be updated by adding the weights of the lower level nodes.
After the propagation, we remove the end node and the edges which connect to it from the call graph.
If there are loops in the call graph, we merge the functions in the loop into one node and remove the branch edge to generate a new end node.
We repeat the process to propagate the weights and remove the end nodes until there are no more nodes in the graph.

\noindent
\textbf{Modified Quality Metric}
Besides adding in the volume size adjustment, we also change the metric from measuring the indirect graph to directed graph since the function calls have directions (from caller to callee function). 
Therefore, when calculating the term $-k_ik_j/2m$ of Equation~\ref{eq_1}~\cite{arenas2007size}, we modify it to incorporate the direction information. Specifically, we only measure the out-degree of the parent node and in-degree of the child node so that we cannot avoid the noise from other irrelevant call edges.
The directed graph model quality metric with volume adjustment is calculated according to the Equation~\ref{eq_2}, 

\begin{equation} 
    Q = \frac {1}{2W}\sum_{i,j}[w_{ij} - \frac{k_i^{out} k_j^{in}}{2W}]\delta(C_i, C_j) \label{eq_2} 
\end{equation}

\noindent
where $w_{ij}$ represents the weight of the edge between function $i$ and $j$, which has the value equal to the function volume weight of $j$. 
$W$ denotes the sum of all the weight for each of the edges in the graph, $k_i^{out}$ and $k_j^{in}$ specify the weighted out-degree of node i and the weighted in-degree of node j, the rest of the notations are the same as Equation~\ref{eq_1}.
With the modified quality score, the function with a large volume will be more likely to be grouped first, since grouping them will output a higher quality score due to their higher weights. 
Therefore, the resulting modules are more coherent than the modules generated by treating all the functions equally. 

\subsubsection{Modularization Algorithm}\label{sec_module_algorithm}
Based on the proposed module quality score, we start to group functions in the program to generate modules. We regard each function as an individual cluster and repeatedly combine two clusters using the fast unfolding algorithm while maximizing the overall quality score. 
Moreover, to make the generated modules more intuitive, we add in two biases to guide the modularization process.

\noindent
\textbf{Fast Unfolding Louvain Algorithm.}
To boost the modularization speed, we choose fast unfolding Louvain~\cite{blondel2008fast}, which is a greedy optimization algorithm, to guide the grouping process. 
The algorithm is adapted to optimize the $Q$ in Equation~\ref{eq_2}.
The modified Louvain algorithm works as follows. 
First, it assigns each node in the network to an individual module. 
Then, it tries to merge any module $r$ with its neighbor module $s$.
The merging will change the module quality by $\Delta Q$ in Equation~\ref{eq_wddq}.

\begin{equation} 
    \Delta Q_{r,s} = e_{r,s}^{in} + e_{r,s}^{out} + e_{s,r}^{in} + e_{s,r}^{out} - 2 * (a_r^{in}*a_s^{in} + a_r^{out}*a_s^{out})\label{eq_wddq} 
\end{equation}
where:
\begin{equation}\label{eq_e} 
    \begin{aligned}
    & e_{r,s}^{in}  = \sum_{i \in r} \sum_{j \in s} \frac{k_i^{in}k_j^{out}}{2W}; 
    & e_{r,s}^{out} = \sum_{i \in r} \sum_{j \in s} \frac{k_i^{out}k_j^{in}}{2W}
    \end{aligned}
\end{equation}
\begin{equation} 
    \begin{aligned}\label{eq_a} 
    & a_r^{in}  = \sum_s e_{s, r} \delta(r, s);
    & a_r^{out} = \sum_s e_{r, s} \delta(r, s)
    \end{aligned}
\end{equation}

\noindent
where the Equation~\ref{eq_wddq}, \ref{eq_e} and \ref{eq_a} can be derived from the previous work~\cite{newman2004fast, arenas2007size}. The notations are the same as Equation~\ref{eq_2}.
The algorithm will merge the community $r$ and $s$, if the merging increases the overall module quality score the most. The algorithm will repeat the same step to greedily merge the nodes until there is no more merging operation could be applied.
The core mechanism of \textit{Fast Unfolding} is the calculation of the change to the global Modularity Quality ($\Delta Q$) for each merging operation. 
To give higher priorities to the nodes that should be firstly clustered according to experts' experience, we introduce two biases to the $\Delta Q$.
The modified $\Delta Q$ calculation is as follows:
\begin{equation} 
    \Delta Q = \Delta Q^{\prime} \times B_l \times B_{e} \label{eq_dq} 
    \end{equation}
where $\Delta Q^{\prime}$ is the basic $\Delta Q$ calculated in Equation~\ref{eq_wddq}.
The $B_l$ and $B_e$ are locality and entry-limit bias introduced to guide the modularization procedures.



\noindent
\textbf{Locality Bias.}
During program development, functions that are designed to perform the same task are likely to be placed together (e.g. in the same source file).
As a result, after being compiled into binary executable, these functions will be placed one after another continuously. With this heuristics, \tool introduces the locality bias to the modularization algorithm. 
The key idea is that we expect to group functions which are close to each other since they have a higher chance to perform the same task.
To achieve this, each function is assigned with an indexing number based on its location sequence in the binary.
Consequently, each module will have an average value of the function indexing.
Then, we define the \textit{dispersion scope} $DS$ of a module as the summation of the distances from each of the functions indexing to the average value.
When merging the two modules, we can update the new values of the average indexing and the $DS$. 
We limit the maximum $DS$ to be the number of functions in the entire program divided by 100. 
If the new $DS$ exceeds the limit, the merging algorithm will be discouraged by 100\% to combine the two modules.
Last, we scale the encouragement and discouragement to the range [0, 3], naming it $B_l$ as the first bias to $\Delta Q$. 
In Equation~\ref{eq_dq}, the $Q^{\prime}$ will be expanded by the $B_l$ from 0 to 300\%.
In this way, we add in the bias to let the algorithm consider the nearer functions first rather than reaching to functions that are very far away.

\noindent
\textbf{Module Entry Limit Bias.}
According to the Single-Responsibility Principle~\cite{single_responsibility_principle}, each method or module should have a single functionality, which should be encapsulated by it. 
We would like the module to have limited entries to ensure the single and encapsulated functionality.
Therefore, we introduce an entry bias $B_{e}$ to during the modularization.
In this work, the module entry is defined as a function node that only has its caller functions outside the module.
The Entry Quality (EQ) score is the number of entries of a particular module.
When calculating the $\Delta Q_{r,s}$ combining module $r$ and module $s$ together, the $\Delta EQ_{rs}$ is defined as the difference between the $EQ$ of the new module and the average value of $EQ_r$ plus $EQ_s$.
After having $EQ$, we calculate the bias $\Delta B_{e}$ according to Equation~\ref{eq_eq_bias}. The $\Delta B_e$ will encourage to merge modules that could decrease the number of entries, and in otherwise discourage to them.
\begin{equation} 
    \Delta B_{e} = 2^{-\Delta EQ} \label{eq_eq_bias} 
\end{equation}

\subsection{Third-Party Library Detection}
\begin{figure}[t]
\centering
\includegraphics[width=\linewidth]{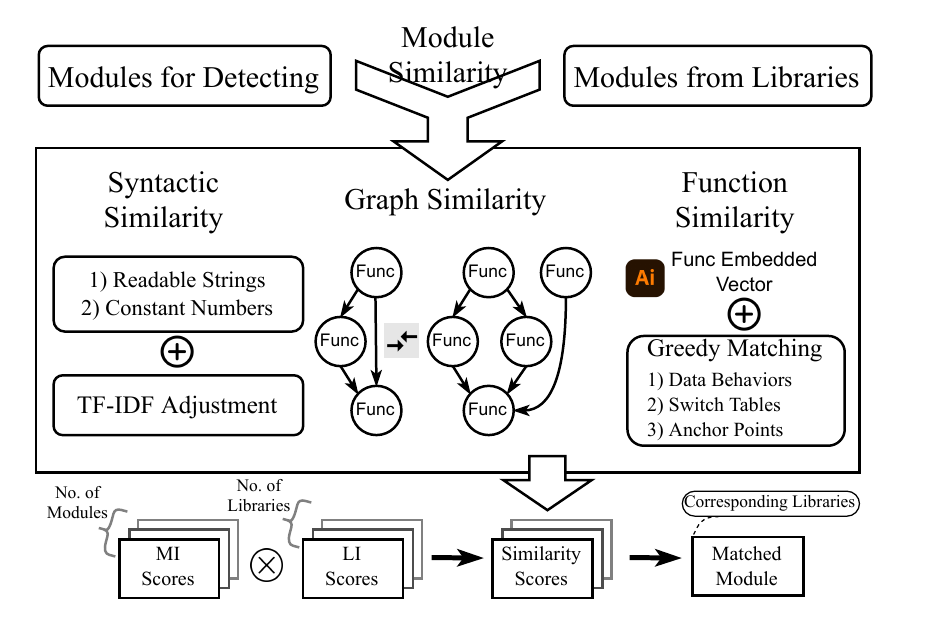}
\caption{Module Matching Overview}
\label{fig:MMO}
\vspace{-5pt}
\end{figure}

After modularizing the program and the TPLs, we propose the similarity measurement algorithm to match the modules based on syntactic and semantic features and detect the TPLs in the program. Figure~\ref{fig:MMO} shows the overview of the TPLs detection procedure via module matching.

\subsubsection{Module Similarity Measurement}\label{module_similarity_measurements}

\noindent
\textbf{Syntactic Features.} Inspired by syntactic feature based library detection works, we incorporate similar features in our module similarity measurement. Specifically, we use the strings literal and constant numbers as the syntactic features. 
String literal is the most significant feature since it usually has unique values, which can be easily distinguished. 
If two functions in two modules have the same string literal, they have a high chance to be the same function. 
However, there are only a small portion of functions which have string literals. Therefore, strings can only help us to accurately match some of the functions and modules. 
Compared to string literal, the constants will have less uniqueness.
For example, we can detect a constant 0, which is used in the branching comparison. 
Meanwhile, constant 0 can be also used to free the memory space. 
Therefore, this kind of constant may not carry useful information for similarity measurement. 
To address it, we adopt the TF-IDF~\cite{ref_tfidf} algorithm to assign more weights to more unique constants, which usually appear less frequently in the module than the rest. 

\noindent
\textbf{Graph Similarity Features.} 
The module consists of functions which call each other to form a call graph. We use propagation graph kernel~\cite{neumann2016propagation} algorithm to measure the call graph similarity.
The algorithm tries to measure the graph and sub-graph structure similarity between two graphs. 
For more fine-grained features, such as each edge of the call graph, we adopt the edge embedding method from RouAlign~\cite{yang2020roualign} to measure the edge similarity in the topology. 
RouAlign promotes a robust way to embed features of function call graphs
With the method, the edges of a particular module could be embedded into vectors. 
And then we could figure out which part of the graph is similar by vector searching, which is time efficient and scalable.

\noindent
\textbf{Function Similarity Features.} These features measure the similarity between functions in the modules.
Since a module consists of multiple functions, the score will be aggregated to measure the module similarity.
To calculate the score, we need to address two problems. 
First, given two functions, how to measure their similarity. Second, how to choose the two functions from the two modules to compare with. 
For the first problem, we leverage a state-of-the-art binary function matching tool Gemini~\cite{gemini2017} to produce a similarity score between two given functions. 
The main idea of Gemini is to embed the function control flow graph into a vector and calculate the vector differences to determine the function similarity. 
Based on our experiment, Gemini has a relatively good performance which can save the time in the feature generation step.

A module may consist of functions with different functionalities.
For example, a module may have functions to perform the core operation, functions to do the error handling, and functions to communicate with other modules. Therefore, we would like to compare functions with similar functionality rather than the ones with different functionalities, which will give a low similarity score.
Moreover, since each module will consist of multiple functions, calculating the pairwise function similarity takes time. Therefore, for the second problem, we adopt a drill-down approach to select function pairs.
As discovered in~\cite{karande2018bcd}, similar functions usually use a common group of data; or they will be referred to by the same pointers. 
Therefore, to selectively measure the similarity, we identify two types of anchor points within the modules to help us to locate functions that are likely to have the same functionalities.
First, in one module, if we detect multiple functions accessing the data in the same memory space, we will mark it as the anchor point (type 1); and we try to detect the similar anchor point in other modules and measure the similarity among the related functions. 
Second, we accessing the dispatch table in the module if it exists. 
The dispatch table is a table of pointers or memory addresses referring to the functions.
We will use these functions as the anchor point (type 2).
We will compare the similarity among the functions that belong to the same type of anchor points.


\subsubsection{TPL Detection}

\tool performs TPL detection by checking whether a module from the target program could be matched to any of the modules in the signature TPLs.
For each module in the target program, \tool matches it against all the modules generated in the signature TPL database by summarizing the similarities between each feature discussed in Section~\ref{module_similarity_measurements}.
\tool ranks the candidate modules by the similarity score and selects the modules with high and distinguishable similarity.

However, the matching result may contain false positives due to the following reasons:
First, some of the libraries may contain similar modules. It is difficult to distinguish from which library the module comes. This will happen especially when the modules are small in size, which will consist of simple structures with few functions.
Second, the TPLs are in different sizes, which will bring unfairness during the matching.
For example, \verb|libbz2| library has only 5 modules with 81 functions, while \verb|libcrypto| library has over 186 modules with 6559 functions. Therefore, if \tool detects a module of library \verb|libbz2|, we may have high confidence that the library is reused in the program. On the contrary, detecting only one module of library \verb|libcrypto| may suggest that it is a false alarm. 

To further improve the accuracy, we adopt two adjustments.
First, we introduce the Module Importance (MI) score to select the modules which are considered to be more important.
In the heuristics, we believe that the bigger the module size, the more important the module would be.
It is because that bigger modules tend to have more unique structures which may not be miss-matched with other modules. 
Therefore, MI is specified in Equation~\ref{eq_module_weight}, where $|m_k|$ denotes the total functions in the $k$-th module, $n$ is the total number of modules.
Second, for a library, its importance ought to have positive correlations with the reference frequency, and negative correlation with the number of the modules that it contains. 
The more frequently one library is needed by other binaries, and the less number of modules the library has, the more important it should be if its modules are detected in the program. 
The Equation~\ref{eq_inverse_library_weight} shows the Library Importance (LI) for library $h$, where the $|l_h|$ denotes the number of modules in the $h$-th library, the $\nu(l_h)$ denotes the times the library $l_h$ is referred to.
It is difficult to determine whether a module is used in the detected binary, but the module usage frequency could be approximated by the library usage frequency.
With this assumption, we give the Matching Confidence (MC) by Equation~\ref{eq_MWILW} to the module $k$ of the library $h$. A higher MC score means the more creditable the detection on the module.
Finally, we combine the similarity scores in Section~\ref{module_similarity_measurements} with the MC to give the final results of the TPL detection.

\begin{equation} 
    MI_k = \frac {|m_k|} {\sum_i^n {|m_i|} / n }\label{eq_module_weight}
\end{equation}
\begin{equation}
    LI_h = \frac {\log(\nu(l_h) + 1)} {|l_h|}\label{eq_inverse_library_weight}
\end{equation}
\begin{equation}
    MC_k = MI_k \times LI_h \label{eq_MWILW}
\end{equation}


\section{Evaluation}\label{sec:eva}
In the experiments, we aim to answer the following research questions:

\noindent
\textbf{RQ1}: What is the quality of the modules generated by \tool compared to other program modularization works?

\noindent
\textbf{RQ2}: What is the accuracy of \tool in detecting TPLs in binary programs compared to related works?

\noindent
\textbf{RQ3}: What is the breakdown performance of \tool in modularization and library detection?

\noindent
\textbf{RQ4}: What are the real-world use cases of partial library detection?

\noindent
\textbf{RQ5}: What are other possible applications of program modularization in software engineering and security?

\subsection{Module Quality Evaluation (RQ1)}
\noindent
\textbf{Module Quality Metrics Selection.}
To evaluate the quality of the generated modules by \tool, we have selected 7 metrics from different aspects. First, since the program modularization process is very similar to the community detection process, we choose the commonly used community quality metrics to measure the modules. ~\cite{newman2004finding} promotes the \textit{Orign MQ}, which measures the quality for an unweighted and undirected network. Moreover, since the program call graph is directed and we have assigned weights to the graph, we also selected \textit{Directed MQ}~\cite{arenas2007size} and \textit{Weighted and Directed MQ}~\cite{khan2017network} as the evaluation criteria. 
Second, we have reviewed the source code level program modularization works and selected 2 metrics used in the state-of-the-art tools' evaluation, namely \textit{Bunch MQ}~\cite{mancoridis1999bunch} and \textit{Turbo MQ}~\cite{mamaghani2009clustering, 2018ArchRecovery}. 
The Bunch MQ~\cite{mancoridis1999bunch} is designed to reward the creation of highly cohesive clusters, and to penalize excessive coupling between clusters.
Turbo MQ is a lightweight metric that includes edge weights.
Last, from the program analysis point of view, we would expect that for each module there should be as few entry points as possible. 
Less entry points suggest that the module can be used/called in less different ways, which ensure the module coherence. 
Moreover, we would like the clustering results to be smooth, which means that there should be as few isolated clusters as possible. 
Therefore, we count the average number of \textit{Entries} and the number of \textit{Isolated Clusters} within each module as the last two metrics.

\noindent
\textbf{Related Work Selection.}
We have chosen two algorithms to compared with to evaluate the module quality. First, as far as we have reviewed, \textit{BCD} is the state-of-the-art binary level program modularization tool in the literature. Therefore, we have compared \tool with BCD on the 7 metrics.
Second, the program developer will tend to place functions with similar functionalities into the same file at source code level.
We can regard each of the files as a program module so that the program is modularized naturally during the development.
Usually, this type of program will be compiled into archive files (".a" as suffix), which consists of many object files (".o" as suffix). 
We measure the quality of the modules generated according to the object file boundaries, denoted as \textit{AR Modularization} and compare it with \tool.

\noindent
\textbf{Module Quality Assessment.}
We have selected 106 commonly used binaries compiled by nix~\cite{nix_dolstra2004imposing} and run \tool and BCD on them.
For AR Modularization technique, since not all the binaries are compiled into archive files, we only tested it on 102 system library binaries, which have the archive files.
Table \ref{evaluation_metrics_comparing} shows the average scores for each of the metrics of \tool, BCD and AR Modularization respectively.
In Table1, the first five metrics are Modularity metrics. Among them, four metrics are used in related works\cite{arenas2007size, 2018ArchRecovery, mancoridis1999bunch, newman2004finding}. 
Modularity\cite{arenas2007size} measures the strength of division of a graph network into modules. The last two metrics are heuristic statistical metrics. They measure the readability and reasonableness of the modules.
Generally, our method reaches higher module quality scores than other modularization methods and has less entries and isolated clusters per module.
The only metric that AR Modularization beats \tool is the \textit{Weighted and Directed MQ}. 
It is because that when calculating the metric, the final score will be normalized against the total weights of the program. 
The programs used to measure the quality for AR Modularization tend to have less weights than the programs used to test \tool and BCD. 
Therefore, AR Modularization has a higher score even if its module quality is lower than other tools.

\begin{table}
    \caption{\bfseries The modularization results for several metrics .}\label{evaluation_metrics_comparing}\vspace{-10pt}
    \footnotesize
    \begin{tabular}{p{2cm}<{\centering}|p{1.5cm}<{\centering} p{1.5cm}<{\centering} p{1.5cm}<{\centering}}
        \hline
        {\bfseries Metrics} & {\bfseries \tool} & {\bfseries BCD} & {\bfseries AR modularization} \\
        \hline {\bfseries Orign MQ~\cite{newman2004finding}}   & 0.020299* & 0.006988  & 0.019758 \\
        \hline {\bfseries Directed MQ~\cite{arenas2007size}}   & 0.019193* & 0.005998  & 0.011387 \\
        \hline {\bfseries Weighted and Directed MQ}           & 0.029362  & 0.016864  & 0.040163*\\
        \hline {\bfseries Bunch MQ~\cite{mancoridis1999bunch}} & 0.007333* & 0.001206  & 0.000403 \\
        \hline {\bfseries Turbo MQ~\cite{2018ArchRecovery}}    & 0.553336* & 0.148786  & 0.045623 \\
        \hline {\bfseries No. of Entries}                     & 1.819864* & 11.799936 & 5.801478 \\
        \hline {\bfseries No. of Isolated Clusters}           & 1.000000* & 15.223941 & 5.737548 \\
        \hline 
    \end{tabular}
    \begin{tablenotes}
       \footnotesize
       \item The * denotes that the score is of the best performance out of the three.
    \end{tablenotes}
    \vspace{-0.5cm}
\end{table}

\noindent
\textbf{Human Labeled Modularization Comparison}
We have collaborated with a big software vendor (name anonymized), which has great interest to the software structure understanding. Therefore, it employs software engineering experts to manually modularize a real-world project \textit{Bash}, which is a commonly used program for command processing. We also compare the results of \tool with it. In this experiment, the source code Bash version 4.2.0 has 
2761 functions. 
The experts manually decompose the software into 13 modules.
Then, we compile the source code into binary and apply \tool to generate 198 modules. 

To evaluate the results, we propose a metric to measure the overlapping between the generated modules and the human labelled modules. We select all the functions in one module generated by \tool and count the number of modules that the same set of functions appear in the manually labelled modules. For example, if a generated module contains three functions A, B and C. Function A belongs to labelled module I, while function B and C belong to labelled module II. Therefore, the overlap metric score will be $2/1 = 2$.
The average overlap score for each generated module is \textbf{1.45}, which suggests that the modules generated by \tool have a high overlap ratio with the human labelled modules. 
Therefore, \tool will be a good solution to save the manpower to produce precise modules automatically.

\begin{figure}
     \centering

     \begin{subfigure}[b]{\linewidth}
         \centering
         \includegraphics[width=0.7\linewidth]{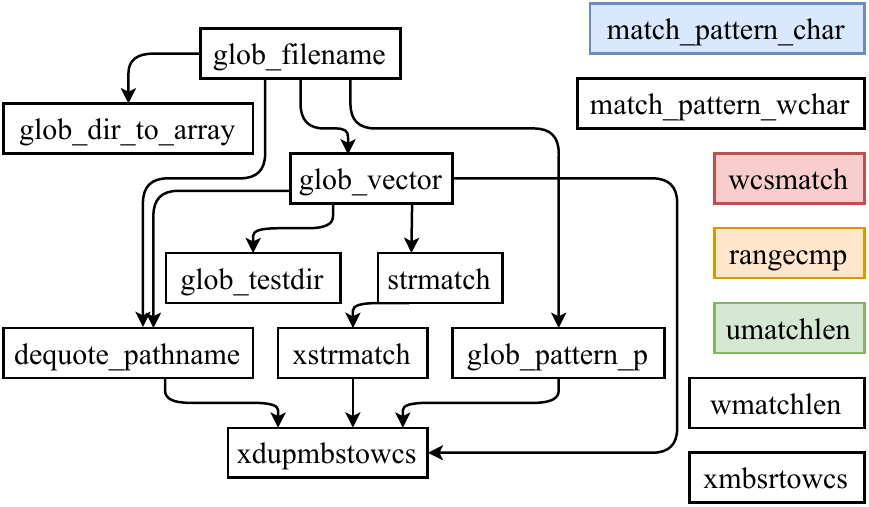}
         \caption{Manually Labelled Module}
     \end{subfigure}
     \hfill
     \begin{subfigure}[b]{0.5\textwidth}
         \centering
         \includegraphics[width=0.75\linewidth]{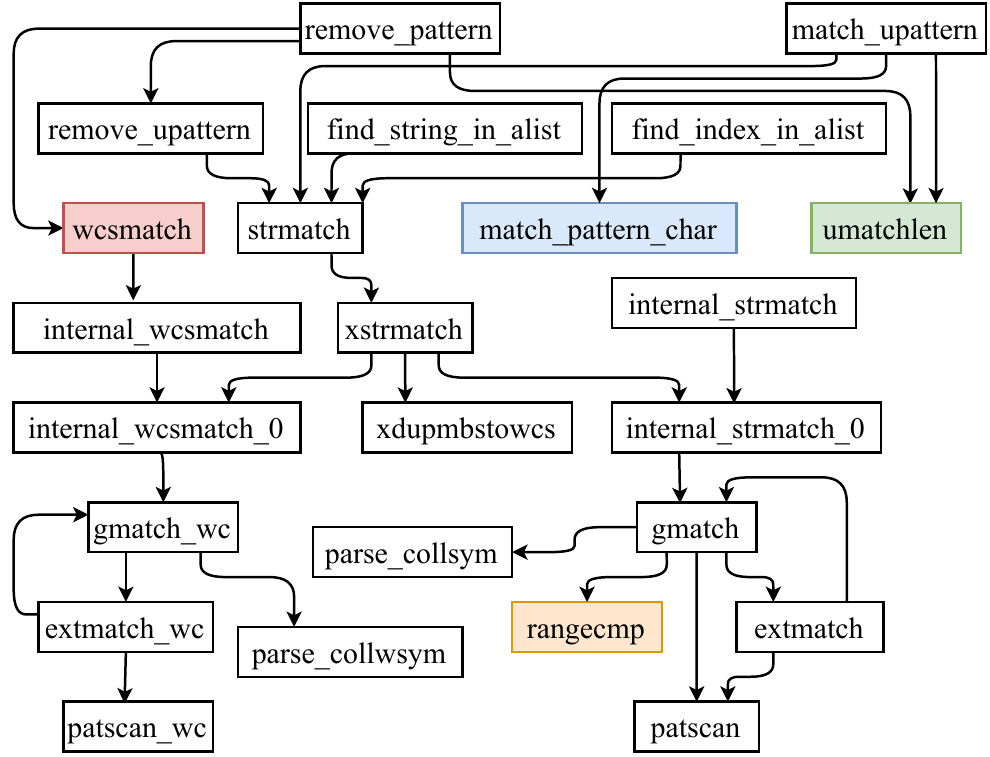}
         \caption{\tool Generated Module}
     \end{subfigure}
\caption{Comparison between Manual and \tool Modularization Results}
\label{fig:module}
\vspace{-0.5cm}
\end{figure}

Moreover, Figure~\ref{fig:module} (a) and (b) shows the concrete example of the modules generated by human experts and by \tool respectively. Since human experts group the source files to form the modules, there may be some isolated functions in each module. As shown in (a), there are 6 isolated functions with 4 being marked in different color boxes. From the names, we know that most of the functions in this module have the similar functionality to process wild-cast strings. For the generated module in Figure~\ref{fig:module} (b), \tool has grouped the 4 isolated functions (marked in the color boxes) into a bigger module with some additional related functions. From the function names, we can notice that most of the functions are with the same functionality, which suggests that \tool has produced a more complete module than the manually labelled approach.

\vspace{2pt}
\begin{tcolorbox}
\textbf{Answering RQ1:}
Compared to the state-of-the-art program modularization work, the average ratio in which \tool outperforms in Modularity Quality(MQ) metrics is 3.53 times.
Moreover, the generated modules are similar to the modules decided by human experts. 
\end{tcolorbox}

\subsection{Library Detection Accuracy Evaluation (RQ2)}

\noindent
\textbf{Binary Program and TPL Data Set.}
We evaluated our tool on two sets of binaries. First, we leverage the package manager, nix~\cite{nix_dolstra2004imposing}, to collect programs with their building dependence on Linux. Nix has provided a service to automatically build binaries with both static-linked and dynamic-linked libraries. We built all available programs under the category "Applications" on nix packages store, and successfully gained 106 binaries with ground truth as the testing data set.
Second, since nix does not guarantee to include all the required libraries in the binaries according to our inspection, to generate the data set with the real ground truth, we manually build a set of binaries on Ubuntu 20.04. Specifically, we choose 7 commonly used programs and build them with statically and dynamically linked TPLs.

To detect the TPLs in the aforementioned binaries, we have also built a TPL database. We have crawled all the 5,278 libraries presented in Ubuntu 20.04. We prune off the duplicate libraries with different architectures and versions and filter out the libraries that cannot be statically linked with the help of \textit{"dpkg"} package manager. 
We order the remaining 795 libraries and choose the top 100 frequently used libraries to form the testing TPL database.

\noindent
\textbf{TPL Detection Tools Comparison.}
To evaluate the TPL detection accuracy of \tool, we choose two state-of-the-art tools, BAT~\cite{hemel2011finding} and OssPolice~\cite{duan2017osspolice}, to compared with.
We run the three tools over the data sets built in the previous step. Since both BAT and OssPolice are designed to detect third-party packages, which contain multiple libraries, we choose to compare the accuracy of both library detection and package detection among the three tools to ensure the fairness.

Table~\ref{real_world_packages} and Table~\ref{known_binaries} show the precision and recall results for the TPL detection tools over nix generated binaries and manually compiled binaries respectively. 
For Table~\ref{real_world_packages}, OssPolice (1) stands for detection results based on our implementation and experiment, whereas OssPolice (2) stands for results claimed in their paper. BAT (1) and BAT (2) have the same meaning. 
From the Table~\ref{real_world_packages}, we can notice that \tool has 83.0\% precision and 73.8\% recall in package detection and 85.6\% precision and 49.6\% recall in TPL detection, which are the highest among the three TPL detection tools.
In Table~\ref{known_binaries}, we list detailed library detection results for the 7 manually crafted Ubuntu binaries. The first two columns present the binary names and the number of TPLs in each of them. The rest of Table~\ref{known_binaries} shows the number of true positives (TP), false positives (FP), and false negatives (FN) for the three tools.
As shown in the table, \tool also achieves the highest results with precision (85.0\%) and recall (65.4\%) on average.


\begin{table}
    \caption{\bfseries TPL Detection on Real-world Programs.}\label{real_world_packages}
    \vspace{-10pt}
    \footnotesize
    \begin{tabular}{p{1.2cm}|p{1cm}<{\centering}|p{1.1cm}<{\centering}|p{1.1cm}<{\centering}|p{0.95cm}<{\centering}|p{0.95cm}<{\centering}}
        \hline ~ & \textbf{\makecell{\tool}} & \textbf{\makecell[c]{OssPolice\\ (1)}} & \textbf{\makecell{OssPolice\\ (2)}} & \textbf{\makecell{BAT(1)}} & \textbf{\makecell{BAT(1)}} \\
        \hline \multicolumn{5}{l}{\textit{Package Detection}}\\
        \hline {\bfseries Precision(\%)} & 83.0 & 83.8 & 82 & 66.1 & 75\\
        \hline {\bfseries Recall(\%)}    & 73.8 & 70.0 & 87 & 65.7 & 61\\
        \hline \multicolumn{6}{l}{\textit{Library Detection}}\\
        \hline {\bfseries Precision(\%)} & 85.6 & 77.8 & /  & 41.4 & / \\
        \hline {\bfseries Recall(\%)}    & 49.6 & 40.2 & /  & 38.7 & / \\
        \hline 
    \end{tabular}
    \vspace{0cm}
\end{table}


\begin{table}
    \caption{\bfseries Partial Library Detection on Ubuntu Binaries.}\label{known_binaries}\vspace{-10pt}
    \footnotesize
        \begin{tabular}{p{0.8cm}<{\centering}|p{0.7cm}<{\centering}|p{0.3cm}<{\centering}|p{0.3cm}<{\centering}|p{0.3cm}<{\centering}|p{0.3cm}<{\centering}|p{0.3cm}<{\centering}|p{0.3cm}<{\centering}|p{0.3cm}<{\centering}|p{0.3cm}<{\centering}|p{0.3cm}<{\centering}}
            \hline
            \multirow{2}*{\bfseries Binary} & \multirow{2}*{\bfseries Libs} & \multicolumn{3}{|c|}{\bfseries \tool} &     \multicolumn{3}{|c|}{\bfseries OssPolice} & \multicolumn{3}{|c}{\bfseries BAT} \\
            \cline{3-11} ~ & {\bfseries Linked} & TP & FP & FN   & TP & FP & FN   & TP & FP & FN \\
            \hline {\bfseries ssldump}  & 2     & 2  & 0  & 0    & 2  & 0  & 0    & 2  & 2  & 0  \\ 
            \hline {\bfseries vim}      & 4     & 2  & 0  & 2    & 1  & 0  & 3    & 1  & 3  & 3  \\ 
            \hline {\bfseries busybox}  & 3     & 1  & 1  & 2    & 1  & 0  & 2    & 1  & 4  & 2  \\ 
            \hline {\bfseries tcpdump}  & 3     & 3  & 0  & 0    & 3  & 0  & 0    & 2  & 1  & 1  \\ 
            \hline {\bfseries openvpn}  & 5     & 4  & 0  & 1    & 3  & 2  & 2    & 3  & 1  & 2  \\ 
            \hline {\bfseries sqlite3}  & 4     & 3  & 1  & 1    & 2  & 2  & 2    & 2  & 2  & 2  \\ 
            \hline {\bfseries openssl}  & 5     & 2  & 1  & 3    & 3  & 2  & 2    & 3  & 1  & 2  \\ 
            \hline
            \hline {\bfseries Total}    & 26    & 17 & 3  & 9    & 15 & 6  & 11   & 14 & 14 & 12 \\ 
            \hline
        \end{tabular}
        \begin{tabular}{p{1cm}<{\centering}|p{1cm}<{\centering}|p{1cm}<{\centering}|p{1cm}<{\centering}|p{1cm}<{\centering}|p{1cm}<{\centering}}
            \multicolumn{6}{c}{\bfseries Performance Summary} \\
            \hline
            \multicolumn{2}{c|}{\bfseries \tool} & \multicolumn{2}{|c|}{\bfseries OssPolice} & \multicolumn{2}{|c}{\bfseries BAT} \\
            \hline Precision & Recall & Precision & Recall & Precision & Recall \\
            \hline 85.0\%    & 65.4\% &  71.4\%   & 57.7\% &  50\%     & 53.8\% \\
            \hline
        \end{tabular}
    \vspace{-0.5cm}
\end{table}


\noindent
\textbf{Discussion.}
In the experiment, most of the binary libraries are partially imported since the modern linkers will only link the used portion of the TPL by default~\cite{levine2001linkers}.
The \tool has better accuracy compared with other tools, because the modules naturally consist of the functions that perform the similar functionality.
When detecting partial usage of the library, the features of modules will keep stable without being demolished.

\noindent
\textbf{FP.}
The bottleneck is caused by the collision of the module features.
There may exist modules with similar structures and functionalities across different libraries. 
The feature extracted from these modules may not be distinguishable enough to separate them.
Therefore, mistakenly matching a module with similar ones in other library signatures will result in the decrease of the precision.
\tool adapts the semantic information into features, which adds in additional feature spaces to increase differences between modules, so that it can produce higher precision in the evaluation.

\noindent
\textbf{FN.}
Since some of the libraries are tiny in size, which only consists of few modules, it is difficult to extract distinguishable features from the limited number of modules. 
Thus, the lack of features in small libraries is the main reason to pull down the overall recall for \tool. Same as many other tools, the \tool will perform better when the versions between the signature library and library in the target function are closer.




\vspace{2pt}
\begin{tcolorbox}
\textbf{Answering RQ2:} Compared to the state-of-the-art TPL detection works, \tool has better on-average precision (85\%) and recall (66\%) on both real-world and manually crafted data set in detecting 100 commonly-used TPLs. 
The semantic module matching and partial library detection capability enable \tool to outperform other works.
\end{tcolorbox}

\subsection{Performance Evaluation (RQ3)}\label{subsection_time_performance}
Table~\ref{time_moduling} gives the average time used to modularize a given program of BCD and \tool. 
Since the time used to modularize the program is proportional to the program size, We divide the testing programs into three size ranges in the experiment.
As shown in the table, in all sizes of binaries, \tool outperforms BCD. 
It is because \tool uses locality scores to guide the rapid modularization.
But in BCD, the locality information is represented as edges between nodes, which makes the graph complicated and slows the process.
\begin{table}
    \caption{\bfseries Program Modularization Time Comparison}\label{time_moduling}\vspace{-10pt}
    \footnotesize
    \begin{tabular}{p{2cm}|p{1.2cm}<{\centering}|p{1.2cm}<{\centering}|p{1.2cm}<{\centering}|p{1.2cm}<{\centering}}
        \hline
        {\bfseries Data Set} & {\bfseries Set A} & {\bfseries Set B} & {\bfseries Set C} & {\bfseries Total} \\ 
        \hline {\bfseries File Size (KB)} & 0 \textasciitilde 100 & 100 \textasciitilde 1000 & > 1000 & 16.4\textasciitilde4413.5\\
        \hline {\bfseries AVG. Size (KB)}      & 61.8  & 297.8 & 2210.2 & 724.8\\
        \hline {\bfseries No. of Binaries}     & 15    &   66  & 25     & 106  \\
        \hline {\bfseries AVG. Func. (\#)}     & 159.6 & 652.1 & 4224.8 & 1425.0\\
        \hline \multicolumn{4}{l}{\bfseries AVG. Modularization Time (seconds)}\\
        \hline {\bfseries \tool}               & 1.4  &  31.7  & 3722.1  & 896.5\\
        \hline {\bfseries BCD}                 & 1.6  &  52.6  & 13650.7 & 3252.7\\
        \hline
        \end{tabular}
\end{table}

Table~\ref{time_detecting} shows the average time used to detect TPLs in given programs. 
Since OssPolice and BAT only use syntactic features, such as strings, which can be indexed, they have better performance than \tool.
\tool extracts semantic features from graphs and measures function similarities, which are mainly unstructured data.
Therefore, we do not have a better way to store and index these features quickly.
We have to load and compare the features one-by-one in the detecting procedure, which lowers the performance. 
A higher accuracy of \tool is guaranteed and is worth the cost of time.
Thus, in practice, we recommend using \tool as a complementary process after syntactic approaches to produce more accurate results.

\vspace{2pt}
\begin{table}
    \caption{\bfseries TPL Detection Time Comparison Time}\label{time_detecting}\vspace{-10pt}
    \footnotesize
    \begin{tabular}{p{3cm}|p{1.2cm}<{\centering}|p{1.2cm}<{\centering}|p{1.2cm}<{\centering}}
        \hline
        {\bfseries Average Detecting Time (s)} & {\bfseries \tool} & {\bfseries OssPolice} & {\bfseries BAT}\\
        \hline {\bfseries Set A (0 \textasciitilde 100 KB)}    & 255.0  & 42.3   & 7.5  \\
        \hline {\bfseries Set B (100 \textasciitilde 1000 KB)} & 915.3  & 81.8   & 32.2 \\
        \hline {\bfseries Set C (> 1000 KB)}                   & 3538.8 & 127.1  & 193.5\\
        \hline {\bfseries Average}                             & 1440.6 & 86.9   & 66.8 \\
        \hline
    \end{tabular}
\end{table}

\begin{tcolorbox}
\textbf{Answering RQ3:} \tool takes on average 897 seconds to modularize binary program which outperforms BCD. However, it costs 1440 seconds to finish the TPL detection, which is slower compared to other approaches. 
\end{tcolorbox}

\subsection{Use Case Study (RQ4)}
\vspace{2pt}
Real-world malware programs usually share only partial codes between variants.
This would be a challenging case to evaluate the partial TPL detection ability of \tool.
We manually collected a family of malware from VirusShare~\cite{virusshare} to perform a use case study.
The malware is from a famous~\cite{wiki_mirai} botnet program family called Mirai, which has been open-sourced since 2016.
It targets at various kinds of networking devices and mutates rapidly.
There are over 100 Mirai variants according to Microsoft collections~\cite{micro_mirai}.
We have selected the original Mirai as the signature to detect the malware appearance in 15 variants submitted from 2016 to 2020 (4 variants in different architectures, 3 variants in the recent year 2020, and 8 other variants). Specifically, we build the malware binary from its source code and add the features into our library database. We regard the malware as a TPL, named \textit{libmirai}.
For each collected malware variants, we detect TPL usage with \tool, BAT and OssPolice.
If \textit{libmirai} is detected in the variants' binaries, we count as a correct malware prediction.

\begin{table}
    \caption{\bfseries Malware Variants Detection }\label{exp_mirai}\vspace{-10pt}
    \footnotesize
    \begin{tabular}{p{2.5cm}|p{1cm}<{\centering}|p{1cm}<{\centering}|p{1cm}<{\centering}|p{1cm}<{\centering}}
        \hline
        {\bfseries Detections} & {\bfseries Total} & {\bfseries \tool} & {\bfseries BAT} & {\bfseries OssPolice}  \\
        \hline {\bfseries Different Architecture}& 4  & 3/4   & 2/4  & 0/1\\
        \hline {\bfseries Variants at 2020}      & 3  & 3/3   & 0/3  & 0/0\\
        \hline {\bfseries Other Versions}        & 8  & 6/8   & 6/8  & 2/4\\
        \hline {\bfseries Total}                 & 15 & 12/15 & 8/15 & 2/5\\
        \hline \multicolumn{5}{l}{\bfseries Summary}\\
        \hline {\bfseries Precision}             & / & 80\%   & 53\% & 40\%\\
        \hline {\bfseries Recall}                & / & 80\%   & 53\% & 13\%\\
        \hline
    \end{tabular}\vspace{-10pt}
\end{table}

Table~\ref{exp_mirai} shows the malware detection results. Overall, our method has the best accuracy in detecting 12 out of 15 malware variants. 
The second row in Table~\ref{exp_mirai} shows that \tool could catch the semantic accurately even across architectures since the semantic based signatures can resist many kinds of modification and mutation. 
The third row shows that \tool is reliable in detecting small partial code reuse, while other tools fail.
BAT uses strings as the signature, which is not stable across variants.
OssPolice is not good at handling binary signatures, leading to the lowest accuracy performance.

\begin{tcolorbox}
\textbf{Answering RQ4:}
\tool has the best malware variant detection accuracy, which suggests that it can detect partial code reuse with the help of matching modules instead of the entire program.
\end{tcolorbox}

\subsection{Applications (RQ5)}\label{sec:app}
In this section, we show other potential applications of the program modularization technique. Besides detecting the TPLs, \tool offers the modularization results for other program analysis works such as reverse engineering and attack surface detection.

\noindent
\textbf{Reverse Engineering with Module Tagging.}
The modules can reveal high level semantic information, which is very helpful for reverse engineering.
As the proof of the concept, we assign tags to the module by extracting the common strings from the function names it contains. Then, we match the module to detect the similar modules in other programs and check if the detected modules share similar tags.
Table~\ref{module_tagging} shows an example of two matched modules with the function names in detail. 
Even though the functions of two modules are different, the tags extracted are similar, which suggests that their functionality at high level are also similar. We manually verify this case to find that both of the two modules try to deal with the connection between the server and the client. Therefore, if we manage to collect different modules with tags as the signatures, we can match the modules in the target program. Then, we can obtain hints about what kind of functionalities the target program has, which is critical in the reverse engineering tasks.


\begin{table}
    \caption{\bfseries Module Tagging Results}\label{module_tagging}\vspace{-10pt}
    \footnotesize
    \begin{tabular}{p{8cm}}
    \hline
        {\bfseries Module.1 Functions} \\
        ssl\_find\_cipher, ssl\_set\_server\_random, ssl\_process\_server\_session\_id, sslx\_print\_certificate, sslx\_print\_certificate, ssl\_process\_client\_key\_exchange, sslx\_print\_dn, decode\_HandshakeType\_ServerKeyExchange,    decode\_HandshakeType\_CertificateVerify,  decode\_HandshakeType\_ClientKeyExchange, decode\_HandshakeType\_Finished, .sprintf, ssl\_decode\_opaque\_array,   decode\_HandshakeType\_ServerHello, decode\_HandshakeType\_Certificate \\
        {\bfseries Module.2 Functions} \\
        tls\_check\_ncp\_cipher\_list, helper\_client\_server, options\_postprocess\_verify\_ce, options\_postprocess, helper\_keepalive, notnull, helper\_tcp\_nodelay, clone\_route\_option\_list, clone\_route\_ipv6\_option\_list, new\_route\_option\_list, init\_key\_type, push\_option, alloc\_connection\_entry, check\_file\_access, rol\_check\_alloc\_0, pre\_pull\_save\_0, .access, check\_file\_access\_chroot, platform\_access, rol\_check\_alloc, ifconfig\_pool\_verify\_range, pre\_pull\_save, cipher\_kt\_get, proto\_is\_net, print\_topology, print\_opt\_route, print\_netmask, print\_str\_int, print\_opt\_route\_gateway, verify\_common\_subnet \\
        \hdashline[1pt/1pt] {\bfseries Common Tags\ } \\
        cipher, type, client, print, server, verify \\
        \hdashline[1pt/1pt] {\bfseries Conclusion in High Level} \\
        Some cryptography handshake between \textit{Server} and \textit{Client}, verifying the identity of the peer.\\
        \hline
    \end{tabular}
\end{table}

\noindent
\textbf{Attack Surface Detection.}
Vulnerability is a special type of program flaw which can lead to security issues. To detect it helps to improve the overall software security. 
According to~\cite{xu2017spain, xiao2020mvp}, functions which contains the vulnerabilities follow certain patterns. 
Therefore, we would like to use the modularization technique to help to identify the attack surface, which aims to determine the modules that are more likely to have vulnerabilities over the others. The security analysis works can benefit from it since they can focus on the vulnerable modules (attack surface) to save time.

To test the attack surface detection ability, we have collected all the CVEs (e.g. commonly known program vulnerabilities) from 5 real-world projects (BinUtils, LibXML2, OpenSSL, FreeType, and Tcpdump). We use \tool to decompose the 5 projects into modules and plot the CVEs to the modules that they belong to.
In the experiment, we focus on the modules, which contain at least one CVE, named Modules-$\alpha$.
Table~\ref{exp_cve} shows the allocation of the CVEs in Modules-$\alpha$ for each of the projects. 
The first few rows show the basic information of the projects and their vulnerabilities. The 8th to 10th rows show the percentage of the number of Modules-$\alpha$ over all modules, the percentage of the number of functions in Modules-$\alpha$ over all functions in the program, and the percentage of the number of CVEs the Modules-$\alpha$ has against all CVEs respectively.

According to the result, we can see a clear indication that Modules-$\alpha$s only account for a small portion of all the modules; but they contain the majority of the CVEs.
For example, in OpenSSL project, 3.7\% modules with 12.4\% functions have 72.5\% CVEs.
Therefore, the modularization technique has the potential to aid the security analysis by providing modules which contain more vulnerabilities and are worthy to be further studied.


\begin{table}
    \caption{\bfseries Distribution of Vulnerabilities in Modules}\label{exp_cve}\vspace{-10pt}
    \footnotesize
    \begin{tabular}{p{1.6cm}|p{0.9cm}<{\centering}|p{0.9cm}<{\centering}|p{0.9cm}<{\centering}|p{0.9cm}<{\centering}|p{0.9cm}<{\centering}}
        \hline 
        ~ & {\bfseries BinUtils} & {\bfseries LibXML2} & {\bfseries OpenSSL} & {\bfseries FreeType} & {\bfseries Tcpdump} \\
        \hline \multicolumn{6}{l}{\bfseries Basic Information.}\\
        \hline {\bfseries Functions}                                 & 1726  & 3108 & 6340 & 1313 & 1266\\
        \hline {\bfseries Modules}                                   & 100   & 267  & 431  & 60   & 91  \\
        \hline \multicolumn{6}{l}{\bfseries Vulnerabilities Information.}\\
        \hline {\bfseries No. of CVEs}                               & 147   & 37   & 70   & 57   & 88  \\
        \hline \multicolumn{6}{l}{\bfseries Distribution of CVEs.}\\
        \hline {\bfseries \% of Modules-$\alpha$}                    & 11.0  & 8.1  & 3.6  & 16.2 & 18.7\\
        \hline {\bfseries \% of Funcs in Modules-$\alpha$}           & 27.6  & 8.3  & 12.4 & 28.2 & 41.3\\
        \hline {\bfseries \% of CVEs in Modules-$\alpha$}            & 85.2  & 66.7 & 72.5 & 89.0 & 78.1\\
        \hline
    \end{tabular}
    \vspace{-0.5cm}
\end{table}

\vspace{2pt}
\begin{tcolorbox}
\textbf{Answering RQ5:} Program modularization has impactful applications in software engineering. Experiments show that it helps to understand the program in reverse engineering and detects attack surfaces in security analysis.
\end{tcolorbox}

\section{Discussion}\label{sec:fw}

\noindent
\textbf{Threats to Validity.}
Our work relies on reasonable modularizations on the program. If the program module semantics changed greatly, our method would lose its effectiveness in matching them. Therefore, two common threats are: 1) Heavy obfuscation on the binaries. 2) Significant semantic changes from the bottom. We acknowledge that these challenges are still difficult to handle and are hot topics in the recent literature.

\noindent
\textbf{Limitations \& Future Works.}
First, as mentioned in Section \ref{subsection_time_performance}, \tool has more overhead compared to other syntactic feature hash matching based approaches. The overhead is mainly introduced by the time to extract features during module matching. One possible solution is to leverage lightweight syntactic matching to filter out obviously irrelevant cases and use \tool to confirm the results in a much smaller candidate space. 

Second, the software researchers have not reached a common consensus about verifying the correctness of the result of binary program modularization.
We have tried our best via proposing our own module metric to measure the quality and evaluating the modules against standard community detection metrics. However, it is difficult to prove that the metrics themselves reflect the real module quality. In the future, we aim to perform an empirical study on the impact of metrics chosen in program modularization since different applications may require different customised metrics for module quality measurement to produce better results. 


Last, the TPL detection is the direct application of program modulization. We believe that this technique has great potential in many other areas. We have evaluated some of the possibilities such as attack surface detection in Section~\ref{sec:app}. In the future, we plan to extend the work to facilitate other analyses in program understanding.

\section{related work}\label{sec:relwork}
In this section, we discuss the related works in the area of program modularization, TPL detection, and code clone detection.

\noindent
\textbf{Program Modularization.}
The program modularization is a helpful technique looking insight into a software system, which is now well developed in source codes analysis. 
Bunch~\cite{mancoridis1999bunch} modularizes source files of the program into clusters by Module Dependency Graph(MDG). 
Following studies~\cite{2018ArchRecovery, mitchell2006automatic, maqbool2007hierarchical, praditwong2010software, huang2016similarity} improve the clustering to realize the automation and the architecture recovery.
Some later studies~\cite{mohammadi2019new, kargar2019multi} can perform modularization more close to human experts.
It is still challenging to modularize a C/C++ binary program and little progress has been made according the newest survey~\cite{alsarhan2020software}. 
C/C++ binaries strip the the structural information of modules after compilation, which in very different from other programs like java applications~\cite{zhan2020automated, ma2016libradar}.
BCD~\cite{karande2018bcd} introduces community detection methods to decompose a binary into modules, and can successfully recover specific C++ classes.
Following studies~\cite{haq2021survey, hamlen2019automated} concludes that the modularization in binary programs is a more semantic approach, and is useful in detecting small pieces of binary code. These works focus on analyzing the program structures with the modularization. Whereas, \tool tries to provide a complete solution to modularize the program and measure the similarity between them.

Many ideas of program modularization come from community detection algorithms. We briefly introduce the algorithms based on the modularity that benefit us.
The original idea was given by Girvan and Newman~\cite{newman2004finding} with an improvement to perform faster at large communities~\cite{newman2004fast}.
Later, Fast Unfolding~\cite{blondel2008fast} was proposed to achieve rapid convergence properties and high modularity output.
After slight migration on the design, variant methods~\cite{arenas2007size, khan2017network} intended for directed and weighted networks were proposed, which are more suitable for the program modularization task.

\noindent
\textbf{TPL Detection.}
TPL detection aims to find the code reuse in software. 
Approaches are proposed to extract the features from source code and match the TPLs in the binary program.
Binary Analysis Tool (BAT)~\cite{hemel2011finding} 
is a representative method based on the usage of constants.
BAT 
extracts the constant values from both sources and binaries, and then utilizes a frequency-based ranking method to identify the presence of third-party packages.
This kind of method is scalable in firmware analysis~\cite{costin2014large, zhang2021capture}.
OSSPolice~\cite{duan2017osspolice} introduces a hierarchical indexing scheme to make better use of the constant and the directory tree of the sources.
BCFinder~\cite{tang2018bcfinder} makes the indexing light weight and makes the detection platform-independent.
OSLDetector~\cite{zhang2020osldetector} builds an internal cloning forest to reduce the efficiency of features duplication between libraries.
B2SFinder~\cite{yuan2019b2sfinder} makes a well study on the features before and after compilation, giving more reliable third-party code detection results.
These methods are designed feature-based rather than semantic-based for efficiency.
Other approaches try to use binary level features to detect TPLs, which are often used in malware analysis.
Native ideas like BinDiff~\cite{flake2004bindiff} and BinSlayer~\cite{bourquin2013binslayer} try to directly match two binaries via graph matching.
LibDX~\cite{tang2020libdx} is a typical tool in TPL detection, with a gene map to overcome the duplication of features, where features are mainly constants for scalability. 
As for java binaries, many methods~\cite{ma2016libradar, li2017libd, zhan2020automated, zhang2019libid} leverage modularized structures to achieve fast and accurate TPL detection.

\noindent
\textbf{Function Level Clone Detection.}
There are also many works identifying function level clones in a binary.
The early methods~\cite{ida_FLIRT} take the bytes code at the function beginning, which is known as IDA FLIRT.
The latter ones~\cite{chandramohan2016bingo, xue2018accurate, hu2018binmatch} extract many internal function features, such as operation codes, control flow graphs~\cite{eschweiler2016discovre}, sequences of basic blocks~\cite{alrabaee2018fossil}, collections of library calls~\cite{BinSim}, symbol execution constraints~\cite{SimByML2018}, and simulate results~\cite{2020DeepBinDiff, pewny2015cross}.
Recently, the state-of-the-arts works~\cite{gemini2017, zuo48neural, Ding2019Asm2Vec} utilize machine learning techniques to achieve the automation in features extraction and clones identification. 
Many clone detection methods have been proved useful in realistic tasks, like vulnerable detection~\cite{xu2020patch}.
These works focus on providing function level features. Our work learns from them to propose unique and robust features for program modules.


\section{Conclusion}\label{sec:conclu}
In summary, we propose \tool to detect TPLs in software via semantic module matching. 
With the novel modularization algorithm, it divides the target program and the signature library into fine-grained functionality-based modules. 
Then, it extracts syntactic and semantic features from modules and measures the similarity among them to detect the presence of TPLs.
Experiments show that \tool outperforms other modularization tools with 353\% higher module quality scores, and outperforms the state-of-the-art TPL detection tools with 17\% lesser false positives.
Moreover, the binary level program modularization technique, as the stand-alone method, also has applications such as reverse engineering and attack surface identification, which provides new research opportunities. 


\section{acknowledgement}\label{sec:acknowledgement}
We appreciate all the anonymous reviewers for their invaluable comments and suggestions. 
This research is supported by the Key Laboratory of Network Assessment Technology of Chinese Academy of Sciences, and the Beijing Key Laboratory of Cyber Security. This research is partially funded by the Strategic Pilot Science and Technology Project of the Chinese Academy of Sciences (Category C, DC02040100) and the Natural Science Foundation of China (NO.61802404, NO.61802394).
This research is supported by the Ministry of Education, Singapore under its Academic Research Fund Tier 3 (MOET32020-0004). Any opinions, findings and conclusions or recommendations expressed in this material are those of the author(s) and do not reflect the views of the Ministry of Education, Singapore.
This research is partially supported by the National Research Foundation, Singapore under its the AI Singapore Programme (AISG2-RP-2020-019), the National Research Foundation, Prime Ministers Office, Singapore under its National Cybersecurity R\&D Program (Award No. NRF2018NCR-NCR005-0001),NRF Investigatorship NRFI06-2020-0022-0001,  the National Research Foundation through its National Satellite of Excellence in Trustworthy Software Systems (NSOE-TSS) project under the National Cybersecurity R\&D (NCR) Grant award no. NRF2018NCR-NSOE003-0001. 
This research is partially supported by the NTU-DESAY SV Research Program 2018-0980.

\bibliographystyle{ACM-Reference-Format}
\bibliography{ref}

\end{document}